# Fire smoke dispersion inside and outside of a warehouse building in moderate and strong wind conditions

Wojciech Węgrzyński[1*], Grzegorz Krajewski[1], Grzegorz Kimbar[1], Tomasz Lipecki[2]

[1] Instytut Techniki Budowlanej (ITB), Warszawa, Poland

[2] Lublin University of Technology, Lublin, Poland

* Corresponding author: w.wegrzynski@itb.pl

**Abstract:**

A moderate-sized (8 MW) fire was modelled in a warehouse located in an urban area with diverse architecture, in wind conditions. The investigation includes the effects of neighboring architecture on the wind field, and in consequence on the smoke control performance and the smoke dispersion in the near-field of the building. Overall, 25 CFD simulations were performed with ANSYS Fluent CFD code, for wind velocities of 5 m/s (moderate), 10 m/s (strong) at 12 wind angles each (0°-330°, 30° increment), and at 0 m/s for reference. The building smoke venting system's performance was affected by the wind, reaching 74%-114% and 78%-158% of the reference mass flow at the moderate and strong wind velocities, respectively. Surprisingly, in multiple cases the exhaust flow rate of ventilators was increased, rather than hindered by the wind. We attribute this to the arrangement of inlets and outlets on façades and the resulting pressure difference between the wall and roof openings. The smoke plume was highly dependent on the wind angle, and the type of architecture up- and downwind of the fire. Significant urban canyon effects and large vortices forming behind tall buildings were observed, leading to smoke accumulation in a large distance from the fire.

## 1 Introduction

The consequences of building fires are routinely assessed with advanced numerical methods such as Computational Fluid Dynamics (CFD) in order to determine the safety of building occupants or the building structure. These efforts are usually focused on the building interior. Successful removal of smoke from the building is considered as a primary goal of fire safety engineering, and its further dispersion in the environment is rarely evaluated.

The removal of smoke and heat from warehouse-type buildings is commonly performed with natural ventilation systems, based on so-called 'Natural Smoke and Heat Exhaust Ventilators' (NSHEV). NSHEV systems are made from automatic openings (vents) at the building façades and roof, that enable the hot smoke to escape from the building while allowing fresh air to take its place. The performance of the vents in fires has been described by Tanaka [1]. It is well known that such systems are prone to adverse wind effects [2], and in some circumstances, it may even negatively affect the consequences of the fire in the building [3]. Given these limitations, the wind is included in the analysis through a wind-condition discharge coefficient $C_{vw}$, that describes the NSHEV performance under in a standardized test at 10 m/s wind, at a most onerous angle. In the past, this approach was questioned as it is limited to a single ventilator's performance, whereas the performance of many devices within the same rooftop will vary significantly [4]. Furthermore, roof obstacles' various aerodynamic effects are not taken into account [5], and the wind effects vary depending on its velocity [6]. The effects of moderate winds and the extent to which the building's surroundings will influence the flow inside the building is not well understood yet. As shown in this paper, the arbitrarily chosen 10 m/s velocity in many cases was not the most onerous, as the performance of the natural ventilators is an outcome of the building aerodynamics, and not just the device itself.

Furthermore, if one shifts the focus from the building-scale fire safety to the city or region-wide consequences of the fire, it will become apparent that smoke released into the atmosphere through NSHEVs may cause adverse effects also outside the building. The investigation of the fallout in the proximity of Grenfell Tower [7] or New York Times press investigation on the concentration of lead in the proximity of the Notre-Dame Cathedral in Paris [8] are examples of recent research about the near-field contamination caused by large fires. A report issued by the Fire Protection Research Foundation includes examples of 32 environmentally significant fires in modern history, focusing on the built environment [9]. Even though the smoke concentration outside is significantly smaller than inside the building, the number of exposed individuals may be much higher, and thus the resulting risk of exposure may be non-negligible. As the smoke may contaminate the soil, buildings or vegetation, smoke exposure effects may be a long-term threat. Investigation of the smoke dissipation effects may not be also valuable for planning safety systems, civil preparedness, and the organisation of the rescue operations. Literature also contains examples of city wide fire modelling including the predictions of building-to-building fire spread based on physics-based models [10]. The current study attempts in defining the near-field pollution from the fire smoke considering different wind scenarios (direction and velocity). This approach could also improve predictions of models such as one introduced in [10].

In this paper we present results of CFD simulations of a moderate size fire (Heat Release Rate, HRR = 8 MW) in a hypothetical warehouse building located in an densely populated urban area (downtown Warsaw) of diverse architecture. As the building is assumed to be built from non-combustible materials, external fire nor the fire spread beyond the building of origin is not investigated. We investigate the fire's outcomes for 24 wind scenarios, which differ in the wind direction (0° – 330°, with 30° increment) and velocity ($u_{ref}$ = 5 m/s and 10 m/s, further referred to as moderate and strong wind, respectively). In addition, a case with no wind ($u_{ref}$ = 0 m/s ) is investigated in reference. The results are investigated for the building's interior and the near-field (local effects in the terminology used in [9]). This area is limited to the building's neighbourhood, at a distance up to 6 times the height of the tallest building in this region. Approach used in here for modelling the near-field may not be applicable to the assessment of the far-field, in which other than CFD types of models may provide accurate results at considerably reduced computational cost [11]. We also recognize that other approaches may be feasible for smoke spread analyses at different scales [11], however the scope of this work is limited to only CFD (with use of ANSYS Fluent) and the near field.

The primary goal of the proposed research is to quantitatively and qualitatively assess the differences in the smoke and exhaust ventilation system performance and the impact of the released smoke on the neighbouring urban development. For the first goal we determine the individual flow rates on different system components and establish a relation between system performance and pressure/flow at the system inlets and outlets. As a secondary goal of the research, we identify the shape of the smoke plume, identify the number of affected buildings and the smoke concentrations at different distances from the fire, in different wind conditions.

## 2 State-of-the-art summary

### 2.1 Natural smoke and heat exhaust ventilation

Natural smoke and heat exhaust ventilation, also referred to as the natural smoke control, is a venting action in which smoke's natural buoyancy is used to push it outside the building through automated openings in the roof or in the upper parts of the façades. The amount of smoke removed from the building is compensated with the fresh air supplied into the building through openings in its lower parts. If executed correctly, the smoke produced in a fire should be only in the 'upper-layer' underneath the ceiling, whilst the bottom part of the building interior remains smoke-free. This is a critical condition for the safety of evacuation and rescue operations. Besides maintaining tenable evacuation routes, the NSHEV removes a substantial amount of heat and smoke, preventing its accumulation inside the building and reducing the thermal exposure of the building structure, installations and occupants. The origins of modern NSHEV systems are traced to the landmark work of Thomas [12], which did lead to the development of modern standards such as NFPA 204 [13]. The development of NSHEV systems was described thoroughly in a review paper [2], and particular challenges related to the design of NSHEVs and their performance in wind conditions in [5]. Other notable studies on NSHEV performance in wind conditions include [3,6,14,15].

### 2.2 Wind and fire coupled modelling for building fires and outdoor fires

Wind can affect the fire and smoke propagation in numerous ways. A wide range of approaches and models used in the prediction of outdoor pollutants is given in [11]. A review of global testing methodologies related to the vulnerabilities of buildings from large outdoor in the built environment is presented in recent ISO/TR 24188:2022 [16]. In the current paper we focus primarily on the near field modelling with the use of CFD and the exposure to smoke produced in an indoor fire. The CFD approach used herein is also commonly used in similar studies focused on pollutant dispersion in various environments. An in-depth literature review of wind and fire coupled numerical analyses were given in [2]. Numerical analyses are used, among others, for the design of more efficient smoke removal from buildings or road tunnels, prediction of smoke flow inside and outside buildings, prediction of smoke production in large outdoor fires and smoke dispersion in both urban and rural environments.

The CFD was commonly used to investigate the wind effects on smoke control systems. An overview of early developments in numerical modelling of fire and wind was performed by [17], who also used CFD to evaluate the performance of natural smoke control in wind conditions. It was observed that high wind velocities could lead to the blockage of fire vents, or even generate reversed flow. Non-uniform pressure distribution as an effect of wind action was studied by Poreh and Trebukov [14], who as the first identified that the probability of wind velocity sufficient to influence the smoke venting was up to 25%. The influence of wind on the performance of natural smoke control systems was investigated by Meroney [2], who found that the wind can modify the flows through external doors and windows and distort the thermal plume of smoke. Similar observations were formed in the current study. In [18] and [4] it was observed how wind acting from different directions can change the mass flow rates at different elements of the smoke exhaust system. This approach was used as a tool for quantitative analysis of the system performance in the current study. In a group of papers [6,19–21] it was identified, that the pressure acting on building facades may change the fire performance of different smoke exhaust system components.

The impact of wind on the fire environment was well understood by the firefighters, for whom the so-called wind driven fires are considered as an important threat during firefighting operations. This was accounted in tactical guidelines developed in the Fire Safety Research Institute (former Firefighter Safety Research Institute) [22,23]. The CFD method was also extensively used to understand the wind influence on fires. [24] performed CFD simulations of the wind-driven fire in a single-storey residential building. A rapid change in thermal conditions in the building was associated with a wind flow of the velocity equal to 4.5 m/s, making the firefighters' operation more difficult. The importance of wind-induced airflow in buildings was emphasised by [25].

A summary of ongoing efforts in studying outdoor fires (there referred to as wildfires spreading to communities, so called Wildland-Urban Interface fires) was given in [26]. The need for model development for such threats was already brough in [27]. Large outdoor fire modelling was the subject of a large workshop [28], which however focused on WUI specific models used mainly for fire propagation in wildfires. The use of ANSYS Fluent was not covered in this resource. However, as shown in the next chapter the ANSYS Fluent software package was extensively used in wind and fire numerical modelling in the past. It must be noted that the modelling of WUI fires often focuses on ignition and fire spread conditions (also through so called spotting with firebrands [29,30]) large scale (region – continent) modelling with simplified models [31] or fire forecasting [32], contrary to our interest which is focused purely on the consequences of smoke dispersion in an urban environment. In recent review [11] different approaches to smoke modelling are identified, with CFD being proposed as a tool for near-field modelling (as in the current paper).

## 2.3 Smoke dispersion modelling with ANSYS Fluent

A study on a city scale pollutants dispersion from a 100 MW fire in a tunnel (emitted through a tunnel portal) was presented by [33]. The author used atmospheric dispersion model CALPUFF coupled with TAPM and CALMET prognostic models to develop the three-dimensional meteorological data and, consequently, determine the most onerous CFD analyses. In the chosen scenario (2.3 m/s wind velocity, at the angle of 92° related to a tunnel portal), CFD analysis using RANS realisable k-ε was performed in ANSYS Fluent to determine the concentrations of CO and PM10. The polyhedral mesh domain had dimensions of 2 km x 2 km x 0.5 km. The paper emphasised the benefits of using a puff model (CALPUFF) to choose the worst-case scenario for which more detailed CFD analysis are performed.

Pollutant dispersion in a group of buildings was studied by [34] with use of RANS standard k-ε and LES models in ANSYS Fluent package. The Authors modelled experiments performed in downtown Montreal in a 1:200 wind tunnel scale [35]. The domain dimensions were based on the COST 732 guidelines [36], and one block in each direction of the urban canyon was explicitly modelled. For the standard k-ε model the wind tunnel measurements were used to determine the vertical inlet profiles of wind velocity, k and ε. For LES, a vortex generation method was used to generate a time-dependent velocity field. For two investigated layouts the agreement between simulations and wind-tunnel tests was good, although better for the LES model. It was observed that the flow separation at the sharp edges of the buildings was crucial for the proper simulation of the concentration fields.

A case-study focused on local removal of outdoor particulate matter in the Eindhoven city centre was presented by [37]. The RANS realisable k-ε model was used in ANSYS Fluent (14.0). Firstly, an in-depth validation against the results of wind tunnel experiments was performed. A passive pollutant source was modelled as the term in the advection-diffusion equation. Satisfactory agreement between CFD simulation and experiment was achieved even in a long distance from the source. The whole domain had dimensions of 4.41 km x 3.57 km x 0.6 km. The area outside the high-fidelity model representing Eindhoven's city centre (area of about 5.1 km²) was defined with respective roughness, in order to obtain the correct formation of the boundary layer. The pollutant sources were located in the streets and car parks and were based on vehicle traffic data. Coupling approach was used to combine the release in underground car parks with the simulation of the exterior. The paper's primary interest was to assess the potential of local filters to reduce the traffic-induced fraction of outdoor PM concentrations. This study showed the applicability of the city-scale CFD analysis in determining the pollutant dispersion and coupling of the interior and exterior of the buildings.

Another research was carried out by [38]. The CFD simulation of a dense, heterogeneous district in Nicosia, Cyprus, was performed and validated using high-resolution data obtained from in-situ measurements. The pollutant dispersion was not modelled explicitly, and the use of the Boussinesq approximation considered the flow buoyancy. Simulations were performed with ANSYS Fluent (16.0) and 3D Unsteady RANS realisable k-ε model. The resulting air temperatures were compared with thermal images of the city and were predicted with satisfactory accuracy. The largest discrepancies were found for the areas with materials not explicitly modelled (e.g. metal covered roofs) and in narrow street canyons. This study illustrates the potential of URANS approach in modelling city-scale buoyant flows. Other relevant recent studies that illustrate the use of RANS to predict city-scale flows are [39] and [40].

Emissions from a fire of a tanker vehicle in an urban area were investigated by [41], using Lagrangian particles within a CFD model. The analysis was focused on time-averaged concentrations of pollutants and their aerial distribution. The dispersion of a hazardous gas emitted from a point source placed in the turbulent flow and located on the ground in Tokyo was simulated with LES and validated with wind tunnel measurements [42]. The results were additionally compared with RANS standard k-ε simulations. It was found that average concentrations in low-velocity wind areas were higher, whereas peak concentrations were higher in high-velocity wind areas.

## 2.4 Fire scenario and assessment criteria

The investigation of the smoke release consequences in an urban environment requires the definition of two key elements: the hazard (source of fire) and the assessment criteria. In the presented analysis, the fire source with HRR of 8.00 MW was adopted. This fire represents moderate-size fire in a small warehouse-type occupation building, with packed commodities stacked to a height not exceeding 4 m. In the Polish law system (relevant for the case study location), the building of the modelled size with standard occupancy would not require sprinkler protection. For buildings with higher risk, sprinklers may be obligatory and assuming their correct operation, the resulting size of fire should be smaller. Finally, larger fires with HRR of 10 – 100 MW and more can create stronger thermal plumes, that affect significantly larger areas further away from the near-field investigated in this study. In summary, the chosen HRR of 8.00 MW is representative for a broad range of warehouse fires and appropriate for the limits of the numerical model used in this assessment. In our case because of the size of the fire and assumption that the building is built with non-combustible materials (according to local law regulations) we exclude the external fire spread or building-to-building fire spread from this analysis. This assumption is confirmed by the result analysis, in which the smoke produced has temperature insufficient to cause ignition of a secondary building. For simplicity, any firebrand generation or transport is not considered in this paper, and the reader is kindly referred to other studies in the literature on this subject [10,29].

The development of a fire with 8.00 MW size takes approximately 7 to 14 minutes (based on "fast" and "medium" fire classification after [13]). The evacuation of the warehouse building equipped with automatic fire alarm can be estimated between 4 – 15 minutes, depending on the technical solution used to signal the alarm and the pre-evacuation time of the building occupants [43]. Firefighters' time to arrive and begin the firefighting operations within a large city can be estimated between 10 – 15 minutes. This means, that all of the processes (fire development, evacuation, start of the firefighting operations) take place in a similar time scale, and can be assessed from the results of 20 minutes transient simulations, as used in this study.

The assessment criteria for the safety within the building are the smoke concentration (visibility > 10 m), temperature (< 60°C) at height of 1.80 m above the floor, and the upper smoke layer temperature (average < 200°C) [44]. The smoke control system's performance is evaluated based on the mass and volumetric flow rate through the ventilators, and relative value compared to the reference case (no-wind conditions).

There is no universal definition for the unsafe smoke exposure outside buildings, nor a general agreement on what smoke concentration value should be used as a threshold for the design of emergency response, preparedness etc. An in-depth review of smoke toxicity and various exposure levels is given by e.g.: [45] and [46]. In case of absence of recommendations, the NOAEL (No Observed Adverse Effects Level) concentration can be used. However, for some substances (e.g. carcinogenic) any acute exposure can be potentially dangerous; thus a no-exposure condition may be preferred. One of commonly used criteria to assess the tenability in fires is visibility in smoke. For outdoor aerosols in the air, the threshold value of transmittance is $\tau = 2\%$ [47]. For the constant value of the specific smoke extinction coefficient $\sigma_s$ = 8.7 m²/g, following [48], and the critical transmittance τ = 2%, the mass concentration of smoke that causes such obscuration at viewing distances of 10 m, 100 m and 200 m is 0.045 g/m³, 0.0045 g/m³ and 0.0023 g/m³, respectively. These values are adopted here to form qualitative conclusions for the near-field of the building.

## 3    Methods

### 3.1    Modelling assumptions

25 CFD simulations using commercial code ANSYS Fluent in version R19.2 were performed. In this Chapter we present the main assumptions for the numerical modelling, while details of used submodels, schemes and approaches may be found in review papers [49] and [11]. The details of the numerical schemes of ANSYS Fluent used herein can be found in its technical documentation [50]. None of the fundamental models were modified, and this chapter provides with all the solver settings used by the Authors in the study.

The hypothetical warehouse building had dimensions of 11.70 x 32.50 m and the height of 8.10 m (at the top of the roof of slope of 5°). Walls of the model were created as a thin wall-boundary condition with 1D heat transfer and assumed 10 cm width, with the thermal properties of concrete blocks (ϱ = 2200 kg/m³, k= 1.2 W/mK, $C_P$ = 880 J/kgK). This warehouse building was modelled in a shape of an existing office building in the same location (coordinates N 52.18552, E 21.01758). The warehouse's exterior was modelled as exists. The interior was empty, besides the frustum shaped volume that represented the fire. The building was connected with the external environment with inlets (two large doors modelled as interior walls) and outlets (smoke ventilators connection modelled as interior walls, exterior modelled as thin walls with properties of steel, d = 1 mm, ϱ = 7750 kg/m³, k= 43 W/mK, $C_P$ = 473 J/kgK). The dimensioning of the smoke ventilators was based on the principals described in [13]. 12 natural ventilators (1.00 m x 1.00 m each, with aerodynamic free area of 0,62 m² [51]) was evenly distributed on the building's roof. The approximate sum of the total aerodynamic free area of the vents forming the system was 7.44 m². The air was supplied through two large entrances, first in the north-facing façade (2.50 m x 3.20 m, A = 8.00 m²) and the second in the west-facing façade (2.20 m x 2.72 m, A = 6 m²). The view of the building is shown in Fig. 1.

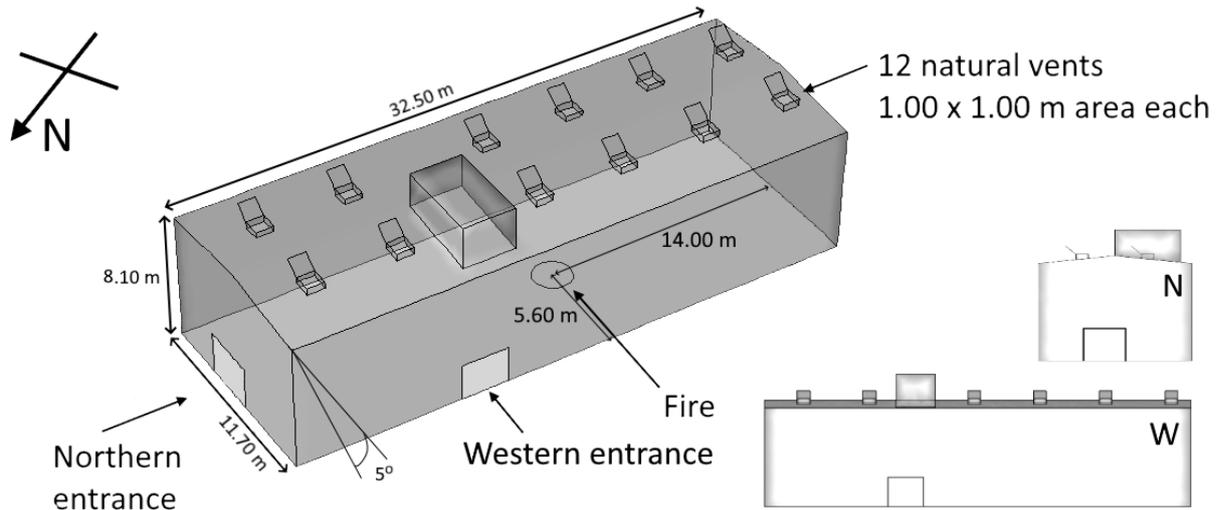

**Figure 1.** The hypothetical warehouse building modelled over a shape of existing building in the same location

The analysis aimed to investigate the smoke dispersion in a near-field of the warehouse, in an urban area with varying architecture. The warehouse neighbours with tall buildings that form an urban street canyon from the East, buildings with a similar height from the West and North and with a large open-field area with vegetation of various height from the South. The height of the tallest building in the analysed area was 60 m. The neighbourhood was modelled based on a terrain map obtained from OpenStreetMap. An overview of the neighbourhood and the numerical model are shown in Fig. 2.

The numerical domain and was subdivided into three areas, as shown in Fig. 2. An area within 30 m from the warehouse building was in the centre of the domain. All buildings were represented with fine details here (detail size > 30 cm). The second part of the domain stretched further for the distance of 300 m. The building models were simplified here, although their shape (footprint) and maximum height were preserved. This area was placed within a cylinder-shaped volume, with a diameter of approx. 750 m. Finally, the cylinder domain was located in a larger box-shaped domain with dimensions of 1 000 m x 1 000 m x 360 m. In this area, the buildings were not modelled explicitly, and the terrain roughness $z_0$ equal to 0.4 [m] was used to represent the effects of an urban environment on the vertical wind velocity profile. The different level of details in the domain was based on the work of [52].

The rotation of the cylinder domain (prior to initialisation of subsequent calculations) was used to account for 12 different wind inflow angles.. For each wind direction, the domain was rotated counterclockwise by the angle of 30°. The initial position of the domain (0°), as well as the subsequent wind directions, are indicated in Fig. 2.

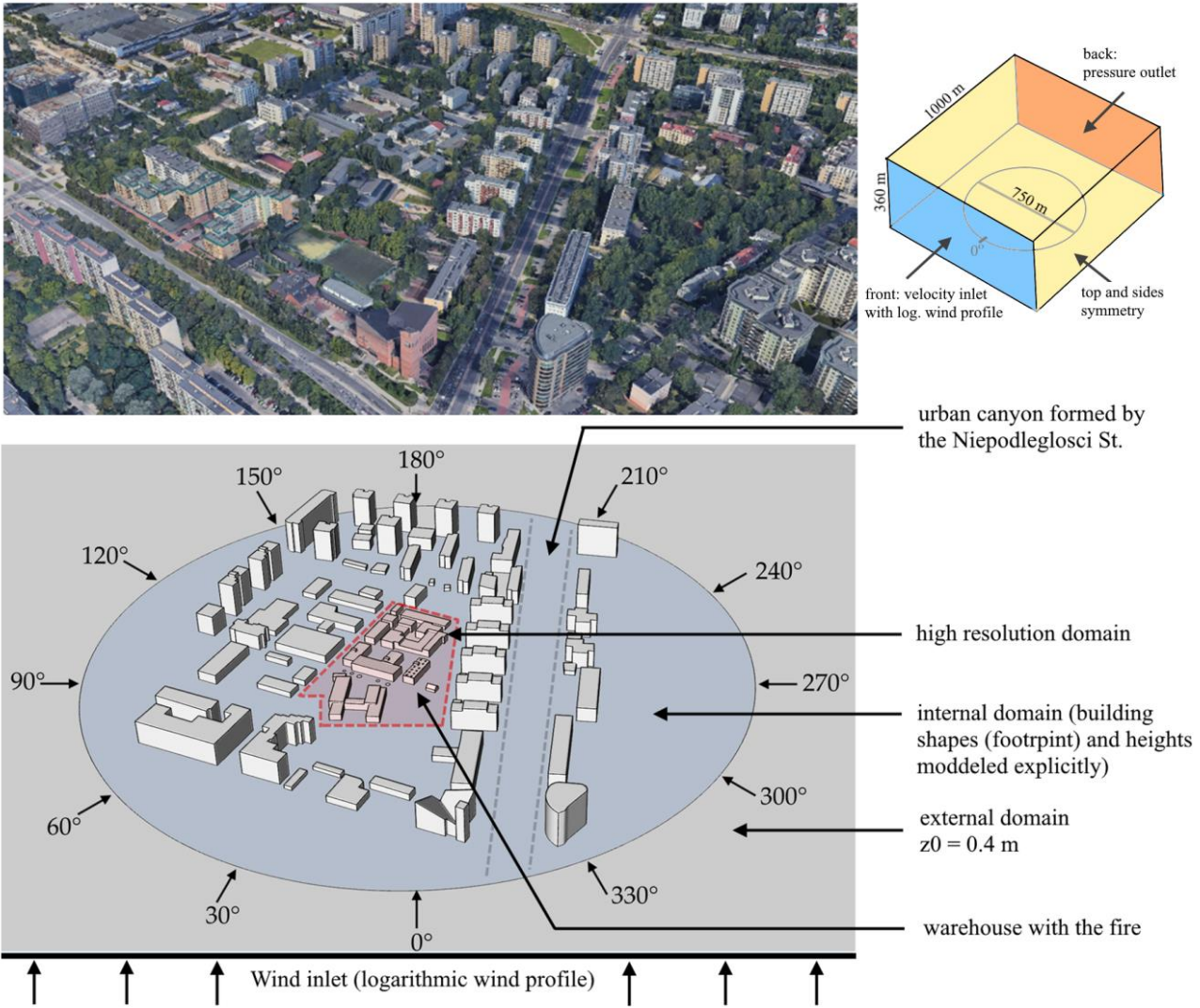

**Figure 2.** The view of the building neighbourhood and the corresponding numerical model. (Source: Google® Earth, own work). Area highlighted in red in the bottom drawign represents the part of the domain with high resolution models of buildings, and area highlighted in blue the part of the domain in which the main building shape (footprint) and their heights were preserved. In the external part of the domain buildings were not modelled explicitly.

Dividing the large domain into three subdomains allowed using different meshing strategies in each part, based on the previous experiences described in [40]. The unstructured tetrahedral mesh was used, primarily due to the ease of meshing the building interior. The mesh inside and on the warehouse's external walls (red building in the middle of high-resolution domain, Fig. 3) had maximum element length of 40 cm, and was reduced to 14 cm on vents, inlets and in the proximity of the fire. The mesh growth rate factor was 1.15 in this area. The mesh on other buildings in the high-resolution part was up to 1 m. In the internal domain, it was up to 2 m. The mesh in the external domain ranged from 2 m to 15 m, and was growing in size with the distance from the internal domain, reaching the maximum size at the domain boundaries. The total number of elements used in the model was approx. 4 000 000. Changes of the mesh are shown in Fig. 4.

The mesh used inside the building (*dx* = 0.14 m to 0.4 m) was considered appropriate for modelling the fire-related flows, based on the estimated value of the characteristic fire diameter D* criterion [53], equation 1.

$$D^* = \left( \frac{\dot{Q}}{\rho_{amb} c_p T_{amb} \sqrt{g}} \right)^{2/5} \quad (1)$$

where $\dot{Q}$ is the Heat Release Rate [kW], $\rho_{amb}$ is the ambient air density [kg/m³], $c_P$ is the specific heat of air at constant pressure [kJ/kg K], $T_{amb}$ is the ambient temperature [K], and g is the Earth gravity [m²/s]. Based on [54], with values of the D*/dx between 4 and 16, the fire plumes were resolved with sufficient accuracy. The volume of the source was 6.65 m³. A higher value here represents a finer mesh. For the fire of 8 MW, D*/dx = 4 and 16 refer to mesh with dx = 0.55 m and 0.14 m, respectively.

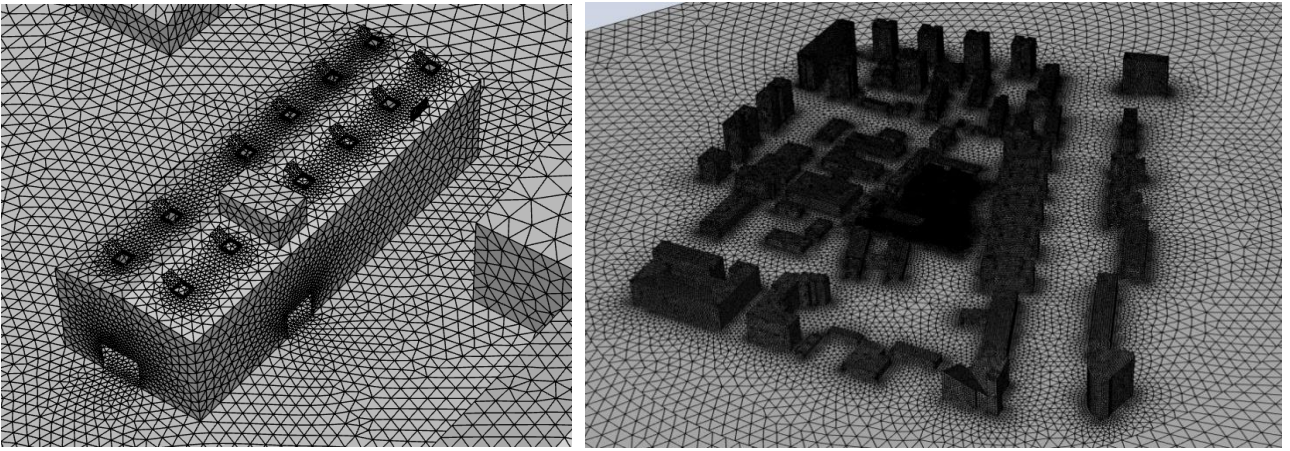

**Figure 3.** Numerical mesh in the high-resolution part of the internal domain (left) and surroundings (right)

The fire was defined through a change of heat, mass and combustion products (constituents of species mixture) implemented as sources to a predefined volume constituting the fire, without explicit combustion modelling, as described in [55]. The fire was located in the central part of the building (Fig. 1), outside of the direct area of influence of the air supply jets coming from the main entrances. The evolution of the HRR was defined with the $\alpha t^2$ relation, with $\alpha$ = 46.70 W/s², commonly known as the "fast" fire [13]. The HRR was limited to 8.00 MW, which was reached by the fire in the 400th second of the analysis. Conservative soot yield value of 0.1 g/g was assumed [56], effective heat of combustion of 25 MJ/kg, which means that the peak source mass flux of soot was 32 g/s.

The wind was introduced at the front boundary of the external domain (velocity inlet boundary condition on Fig. 2) as a logarithmic vertical wind profile, with variable velocity (*u*), turbulent kinetic energy (*k*) and its dissipation rate (*ε*), Eq's 2-4. No additional changes to the wall boundary conditions in the model were introduced, and the wind profile was verified at the edge of the domain. The back side of the domain was modelled as a pressure outlet boundary condition, and the side and top walls as symmetry.

$$u(z) = \frac{u_*}{\kappa} \ln\left( \frac{z + z_0}{z_0} \right) \quad (2)$$

$$k(z) = \frac{u_*^2}{\sqrt{C_\mu}} \quad (3)$$

$$\varepsilon(z) = \frac{u_*^3}{\kappa(z+z_0)} \quad (4)$$

where: $z$ – height [m], $z_0$ – aerodynamic surface roughness length [m], $\kappa$ – von Karman constant [-] (0.40-0.42), $C_\mu$ – model constant [-] (0.09), $u_*$ – friction velocity [m/s].

The wind velocity $u_{ref}$ was set to either 0 m/s, 5 m/s or 10 m/s measured at the reference height $z_{ref}$ = 10 m. These wind scenarios are further referred to as the moderate and strong wind, respectively, The terrain roughness in the external domain was $z_0$ = 0.40 m. The mean roughness height of building walls was 0.01 m.

## 3.2  Solver settings

The simulations were performed with a double-precision 3D solver mode [50], as transient. The mass conservation equation (5) of the CFD model means that the air and smoke can be removed from the model only when it flows out through the domain boundary. The density change in any volume equals the mass flow through its boundaries or the mass introduced through the source term ($\dot{m}_i'''$).

$$\frac{\partial p}{\partial t} + \nabla \cdot \rho \vec{u} = \dot{m}_i''' \quad (5)$$

where: $p$ – pressure, $t$ – time, $\rho$ – density and $u$ – velocity.

The transport of smoke is modelled as with a mixture fraction sub-model (soot defined as one of the species $Y$), and its transport can describe by the equation (6).

$$\frac{\partial(\rho Y_i)}{\partial t} + \nabla \cdot (\rho Y_i \vec{u}) = \nabla \cdot (\rho D_i \nabla Y_i) + \dot{m}_i''' \quad (6)$$

where: $D_i$ – is the dispersion coefficient of i-th species.

The momentum conservation equation (6) is expressing the preservation of the Newton's Second Law of Motion. The forces causing the fluid flow are composed of the pressure field $\nabla p$, tension (tensor $\bar{\bar{\tau}}$), buoyant force $\rho \vec{g}$ and external forces $\vec{F}$. The turbulence was resolved with k-ω SST model (complete model description in [57]) without any modifications to the default constants of the model or its settings. In this case the turbulence modelling approach should be considered as unsteady-RANS.

$$\frac{\partial(\rho \vec{u})}{\partial t} + \nabla \cdot (\rho \vec{u} \vec{u}) = -\nabla p + \nabla \cdot (\bar{\bar{\tau}}) + \rho \vec{g} + \vec{F} \quad (6)$$

Since the simulations performed took into account the difference in the temperature inside and outside the building, the system was modelled taking into account the energy conservation equation (7), which identifies that the enthalpy (8) in any point changes according to the stream of energy flowing into the control volume (in our case the energy source term to the fire source $\dot{q}'''$). Heat can also be delivered as a result of the fluid kinetic energy dissipation as a result of friction ($\varepsilon$), the impact of pressure ($\frac{Dp}{Dt}$) or heat radiation. In fire-safety related applications, the term responsible for the pressure field impact or kinematic energy dissipation is usually neglected.

$$\frac{\partial(\rho h)}{\partial t} + \nabla \cdot (\rho h \vec{u}) = \frac{Dp}{Dt} + \dot{q}''' - \nabla \cdot \vec{q} + \varepsilon \tag{7}$$

$$h = \int_{T_0}^{T} c_p dT \tag{8}$$

The fluid was modelled as a mixture of ideal incompressible gasses. A change in the density can be sufficiently described by a perfect gas equation (9). $M_{avg}$ (10) in equation (9) stands for the averaged molecular weight of the gas mixture ingredients, whose concentration can be identified in any volume using (5).

$$p = \frac{\rho \Re T}{M_{avg}} \tag{9}$$

$$M_{avg} = \frac{1}{\sum \frac{Y_i}{M_i}} \tag{10}$$

Where: $p$ – pressure, $\Re$ - gas constant, $T$ – temperature, $Y_i$ – vol. concentration of i-th species, $M_i$ – molar mass of the i-th species.

The radiative heat transfer was modelled with Discrete Ordinates model [58] (pre-defined 162 discrete angles), and the heat transfer to the walls was modelled as a third-type boundary condition (combination of convection and radiation). Radiative heat transfer coefficient at walls and of the smoke was approximated to ε = 0.6. The heat transfer within the walls was modelled with 1D implementation of the Fourier law. A summary of the essential solver settings is given in Table 1.

**Table 1.** Summary of relevant solver settings for the CFD calculations

| Mathematical models | |
|---|---|
| Solver | pressure-based |
| Turbulent flow sub-model | k-ω SST |
| Time discretisation | unsteady analysis, variable time step = 0.1 – 0.5 s |
| Total length of the simulation | 1200 s |
| Radiation heat-transfer sub-model | discrete ordinates |
| Convective heat-transfer sub-model | based on the Fourier law |
| Computational scheme | PISO |
| Sub-model schemes | all sub-models as second-order |
| Under-relaxation coefficients | ANSYS Fluent defaults |
| **Initial and boundary conditions** | |
| External and supplied air temperature | 20°C |
| Wall temperature (initial) | 20°C |
| Wall roughness height (buildings) | 0.01 m |
| Fluid material | air (incompressible ideal gas) |
| Operating pressure | 101350 Pa |
| Fluid density | 1.205 kg/m$^3$ at 20°C |
| Fire representation | volumetric source of heat and mass [55] |
| Heat Release Rate (peak) | 8.00 MW (achieved in the 400$^{th}$ second of the analysis) |
| Heat Release Rate per unit of volume | 1 200 kW/m$^3$ (at the peak HRR) |
| Fire growth rate coefficient | 46.70 W/s$^2$ (so called fast fire growth) |
| Soot yield | 0.1 kg/kg |
| **Convergence criteria** | |
| Mass, energy, k | 10$^{-4}$ |
| Radiation model, ω | 10$^{-6}$ |

## 3.3 Model validation and best practices

Based on the scale of analysis and phenomena investigated, modelling of wind and fire is similar to modelling microscale (<2 km) of urban climate, which is commonly performed with CFD simulations [59,60]. Introduction of a strong buoyant plume of fire smoke increases the complexity of such analyses, and is inherently burdened with uncertainty. Unfortunately, due to associated cost and complexity no full-scale quantitative validation studies of fires in urban environment exists. Some attempts to model outdoor fires or pollutant dispersion with the same tool (ANSYS Fluent) as used herein was presented in Chapter 2.3. A review of the topic of the dispersion spread of pollutants was published by Lateb et al. [61], and a summary of best practice guidelines in this field was published by Meroney et al. [62]. Furthermore, feasibility of modelling particulate dispersion were demonstrated by Blocken et al. [37] and Luo et al. [63]. A validation experiments for RANS simulations were performed by Nozu and Tamura [42]. More studies related to the modelling fires in environment can be found in [2] and [11]. The cited studies show that urban-scale modelling of the consequences of pollutant release is feasible. The validation of modelling the natural ventilators with CFD approach (same as used in this paper) based on measurements performed in a wind tunnel was presented in [4].

Based on the literature review we have identified, that use of ANSYS Fluent software with k-ω SST turbulence model can be considered as a common approach in both wind and fire engineering, thus providing widely accepted modelling platform for our considerations. As no experimental proof can be provided for the hypothetical warehouse fire presented here, we base the validity of our study on the literature and adopting widely recognized best practices for urban-climate and fire modelling. Recognizing this, we have attempted to minimise the sources of uncertainty, by:

(a) rigorously following the best practice guidelines presented in [49], and the widely accepted guidelines for computational wind engineering [36,64,65];

(b) using the commercial code ANSYS Fluent with a choice of sub-models that has been validated in studies of pollutant dispersion;

(c) creating the numerical domain in accordance to the guidelines [49] and at the same time fulfilling the commonly used criteria for fire models [53];

(d) implementing the experiences from a past study, in which we validated the simulation of the flow through a natural ventilator in wind conditions with results from the wind tunnel tests [5].

## 4   Results

### 4.1   Qualitative and quantitative assessment of the smoke control system performance in wind conditions

The wind direction and velocity had a significant impact on the flow field inside and around the assessed building. The largest differences were observed between cases where the wind was not influenced by surrounding tall buildings (eg. 30°, 90°) and cases where the warehouse was shielded by tall buildings in its near neighbourhood (eg. 240-300°), as visible on Fig 2.

The qualitative assessment of smoke control system performance was based on the vertical temperature profile inside the building, as shown in Fig. 5. These results are presented for cases of 0°, 270° and 330° wind angles, which respectively had the highest (negative), highest (positive) and lowest differences in the mass flow rate on vents in moderate and high wind velocities. In each case the depth of smoke layer was approx. 2 m and the differences in the height of the smoke layer bottom were not significant, and the smoke-free area was well maintained above the floor in all cases. In the case of 0° wind angle and at velocity of 10 m/s the thermal plume was disturbed (skewed), which is discussed further. However, even in this case, there was no significant effect of smoke washout from the plume, leading to negative assessment of the system performance. In all simulations with the wind, the smoke control system met the assessment criteria defined for such systems [13].

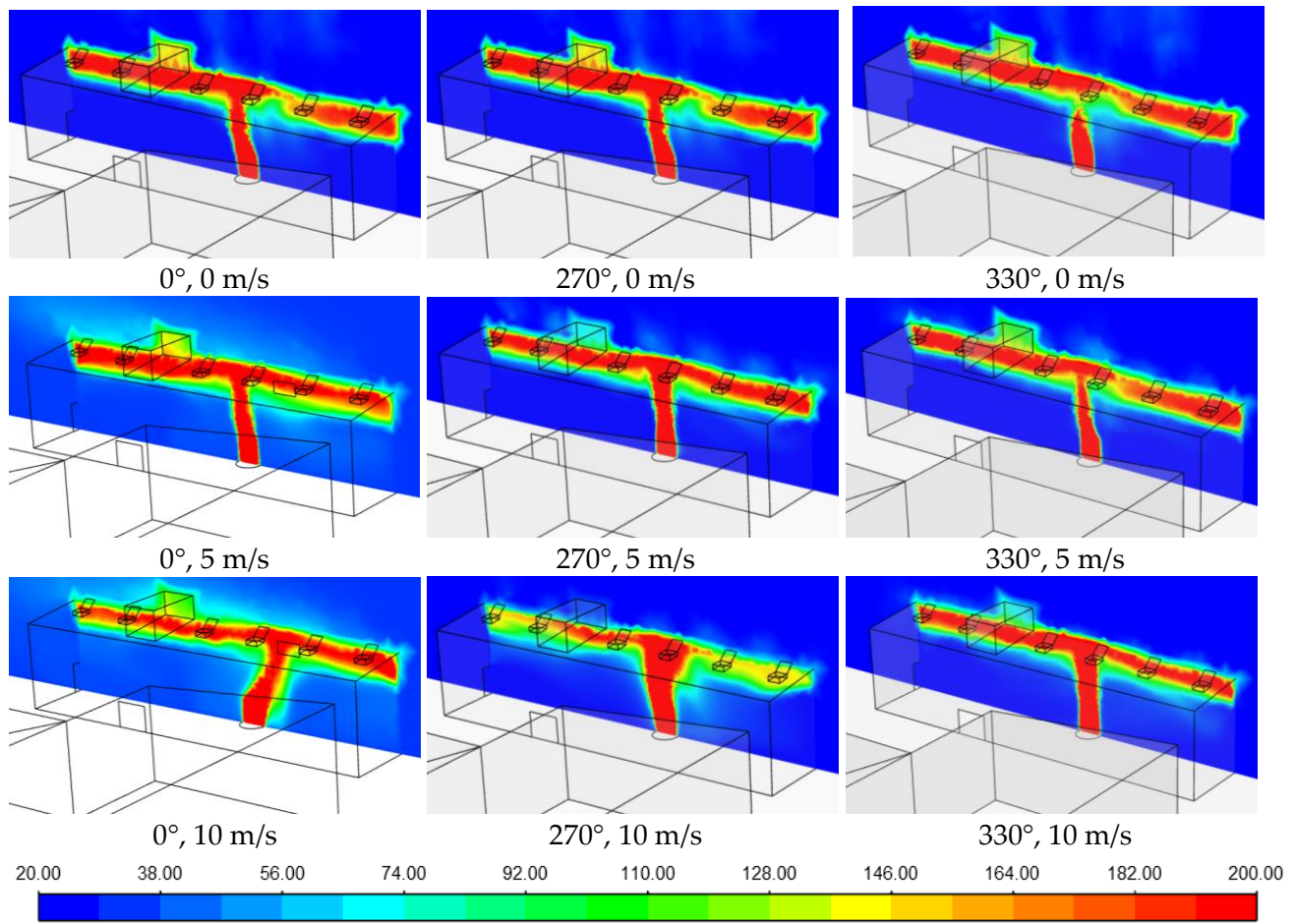

**Figure 4.** Temperature (20 – 200°C) in a cross-section of the building for 0°, 270° and 330° wind angles at 0 m/s, 5 m/s and 10 m/s wind velocities. Plots are from the 10th minute of the CFD analysis.

The smoke venting performance was assessed quantitatively through the estimate of the total mass flow rate through the natural ventilators, which may be considered as a measure of "how much smoke is removed from the building", Table 2. It should be noted that the soot production at peak is 32 g/s, while the amount of smoke is a product of mixing the soot with air in the entraining thermal plume. Thus, the estimation of the "mass of the smoke", here understood as the soot and air mixture, is difficult define as it varies in space and time. As a consequence, we have estimated the baseline performance of the system as the mass flow rate through ventilators at no-wind conditions (27.2 kg/s). The measured mass flow rate of the system changes with the wind angles and velocities. In case of moderate wind ($u_{ref}$ = 5 m/s) the mass flow rate of the system varied from 18.92 kg/s (at 90°) to 30.96 kg/s (at 330°). In case of strong wind ($u_{ref}$ = 10 m/s) the minimum and maximum values were respectively 21.16 kg/s (at 0°) and 42.87 kg/s (at 270°).

For most wind angles the mass flow rate in wind conditions was lower than the rate estimated without wind. Only in 7 out of 24 investigated cases with the wind the average rate was higher than in the reference case (without wind). In general, the highest averaged mass flow rates were calculated for the angles corresponding to the flow from the less heavily built-up area containing street canyon. The largest differences between the mass flow rate values calculated for the same wind inflow angle and 5 m/s and 10 m/s wind were found for 30° and 240° – 270°. In these cases, the flow came from less build-up areas containing low buildings and tall buildings forming street canyon, respectively. The lowest differences were found for angles 300° – 0°. This could be caused by the presence of many high buildings at the leeward side of the warehouse. In the case of high wind velocity, significant differences were found for the air velocity at doors, which is investigated in details further. In Table 2 besides values of mass flow rate and its relation to the reference are also collected other parameters allowing comparison of different inflow cases.

**Table 2.** Mass flow rates at the smoke ventilators, volumetric flow rates at doors and the temperature of the smoke for the analysed cases

| ID | Angle | Mass flow rate through vents | (Mass flow rate) / (Mass flow rate at reference) | Volumetric flow at inlets (doors) | Avg. flow velocity at inlets | Avg. temperature of smoke |
|---|---|---|---|---|---|---|
|  | [°] | [kg/s] | [%] | [m³/s] | [m/s] | [°C] |
| 0 m/s (no-wind) | | | | | | |
| 1 | 0 | 27.2 | 100% | 41.5 | 2.96 | 220 |
| 5 m/s | | | | | | |
| 2 | 0 | 20.11 | 74% | 41.7 | 2.98 | 219 |
| 3 | 30 | 20.13 | 74% | 42.8 | 3.06 | 228 |
| 4 | 60 | 24.29 | 89% | 39.3 | 2.81 | 234 |
| 5 | 90 | 18.92 | 70% | 44.1 | 3.15 | 231 |
| 6 | 120 | 27.97 | 103% | 43.4 | 3.10 | 240 |
| 7 | 150 | 26.04 | 96% | 41.8 | 2.99 | 233 |
| 8 | 180 | 26.81 | 99% | 42 | 3.00 | 199 |
| 9 | 210 | 20.24 | 74% | 40.7 | 2.91 | 208 |
| 10 | 240 | 30.84 | 113% | 44.3 | 3.16 | 198 |
| 11 | 270 | 30.71 | 113% | 43.6 | 3.11 | 225 |
| 12 | 300 | 25.92 | 95% | 40 | 2.86 | 211 |
| 13 | 330 | 30.96 | 114% | 41.1 | 2.94 | 250 |
| 10 m/s | | | | | | |
| 14 | 0 | 21.16 | 78% | 42.3 | 3.02 | 209 |
| 15 | 30 | 30.22 | 111% | 49.3 | 3.52 | 213 |
| 16 | 60 | 21.22 | 78% | 39.3 | 2.81 | 190 |
| 17 | 90 | 23.39 | 86% | 39.3 | 2.81 | 243 |
| 18 | 120 | 22.05 | 81% | 35.5 | 2.54 | 244 |
| 19 | 150 | 21.60 | 79% | 36.4 | 2.60 | 253 |
| 20 | 180 | 23.88 | 88% | 38.4 | 2.74 | 234 |
| 21 | 210 | 27.76 | 102% | 41.8 | 2.99 | 214 |
| 22 | 240 | 38.97 | 143% | 49.1 | 3.51 | 162 |
| 23 | 270 | 42.87 | 158% | 53.8 | 3.84 | 164 |
| 24 | 300 | 28.38 | 104% | 40.8 | 2.91 | 223 |
| 25 | 330 | 30.61 | 113% | 43 | 3.07 | 232 |

## 4.2 The effects of different wind angle and velocity on the airflow patterns outside and inside the building

The differences in the mass flow rate for various wind angles can be associated with the flow relative to the warehouse inlet openings and the differences in the pressure field in the surrounding of the building. The static pressure values at smoke ventilators (average and SD) and inlets are shown in Table 3. Large overpressure at the northern inlet (door 1) was associated with cases with the highest mass flow rate on the smoke ventilators (38.97 kg/s and 42.87 kg/s, at a pressure difference of +8.4 Pa and +11.9 Pa, for 10 m/s wind velocity and 240° and 270° wind inflow angles, respectively). Lower mass flow rates were generally occurring at lower pressure differences, although no clear trend was found in the data. Similarly, the pressure difference at the western inlet (door 2) was not found to impact the flow on the exhaust points significantly.

In the 0° case we have observed the lowest difference in pressure between roof ventilators and inlets to the model, as the flow was parallel to the wall or the opening was located on the leeward side of the building. The pressure pattern was similar in the 330° case (which differs from 0° by only 30°). However, we have observed a jet of air affecting one building inlet directly. In the 270° case the same effect was more apparent, as we have observed a jet of air perpendicular to the largest inlet of the building, and a flow recirculation affecting the second inlet to the building. In this case, the pressure difference between inlets and outlets of the building was the greatest. This is in line with recommendations of [49], where the cases with largest pressure differences between inlets and outlets could be in general associated with the extreme cases for the smoke control in buildings. However, the performance of the ventilation system cannot be predicted entirely by the pressure differences only.

**Table 3.** Static pressure measured at building inlets and outlets for different cases

| ID | Angle | Average static pressure at outlets | Standard deviation of static pressure at outlets | Static pressure at door 1 | Static pressure at door 2 | Pressure Difference (Outlets – Door 1) | Pressure Difference (Outlets – Door 2) |
|---|---|---|---|---|---|---|---|
| | [°] | [Pa] | [Pa] | [Pa] | [Pa] | [Pa] | [Pa] |
| | | | 0 m/s (no-wind, reference case) | | | | |
| 1 | 0 | -1.9 | 0.6 | -2.4 | -2.5 | -0.5 | -0.5 |
| | | | 5 m/s | | | | |
| 2 | 0 | -5.1 | 2 | -5.5 | -5.4 | -0.4 | -0.2 |
| 3 | 30 | -5.4 | 1.7 | -5.3 | -4.8 | 0 | 0.5 |
| 4 | 60 | -3.9 | 2 | -4.6 | -3.7 | -0.7 | 0.2 |
| 5 | 90 | -5.3 | 1.9 | -6.4 | -4.9 | -1.1 | 0.4 |
| 6 | 120 | -6.6 | 2.6 | -7.4 | -6.6 | -0.9 | -0.1 |
| 7 | 150 | -4.4 | 1.6 | -6.1 | -5.3 | -1.7 | -0.9 |
| 8 | 180 | -7.9 | 1.4 | -9.7 | -11.1 | -1.8 | -3.2 |
| 9 | 210 | -7.2 | 0.9 | -7.4 | -7.3 | -0.2 | -0.1 |
| 10 | 240 | -6.4 | 1.1 | -4.8 | -8.3 | 1.6 | -1.9 |
| 11 | 270 | -6.5 | 1.5 | -5.8 | -6.9 | 0.7 | -0.3 |
| 12 | 300 | -5.9 | 0.8 | -7.9 | -7.5 | -1.9 | -1.5 |
| 13 | 330 | -4.1 | 1.7 | -5.6 | -5.5 | -1.5 | -1.4 |
| | | | 10 m/s | | | | |
| 14 | 0 | -12.9 | 6.1 | -11.1 | -12.4 | 1.7 | 0.4 |
| 15 | 30 | -18.5 | 1.7 | -19.4 | -19 | -0.9 | -0.5 |
| 16 | 60 | -10.6 | 3.3 | -15 | -13.7 | -4.5 | -3.2 |
| 17 | 90 | -12.9 | 2.9 | -12 | -8.7 | 0.9 | 4.2 |
| 18 | 120 | -15.2 | 2.3 | -18.6 | -16.8 | -3.4 | -1.5 |
| 19 | 150 | -15.6 | 1.6 | -19 | -17.3 | -3.4 | -1.7 |
| 20 | 180 | -21.4 | 2.1 | -26 | -26.9 | -4.6 | -5.5 |
| 21 | 210 | -18.2 | 1.6 | -18.5 | -16.3 | -0.2 | 1.9 |
| 22 | 240 | -17.3 | 2.4 | -9 | -19.4 | 8.4 | -2 |
| 23 | 270 | -23.9 | 3.6 | -12 | -20.4 | 11.9 | 3.5 |
| 24 | 300 | -19.8 | 4.6 | -16.7 | -16.3 | 3.1 | 3.5 |
| 25 | 330 | -7.2 | 5.4 | -5.9 | -6.9 | 1.3 | 0.3 |

An interesting observation can be made towards the differences in mass flow rate at the same direction of wind but at different velocities, as shown in Table 4. For some wind angles (eg. 30° or

270°) the mass flow rate through vents was 33% and 28% smaller at 5 m/s than at 10 m/s wind. In contrary, for other angles (eg. 120° or 150°) the flow at 5 m/s was 27% and 21% higher than at 10 m/s. It should be noted, that for all of the abovementioned cases the system performed at 89%-103% of the reference no-wind case. We asssociate most of the observed differences between wind velocities with the different pattern of flows forming at the inlet and roof levels, and in consequence different flow at the inlets (as shown in Table 4). These changes are consistent with observed changes of the mass flow rate (with the exception of the 90° wind angle case). For the 90° wind angle we have observed a distinct change in flow patter on door in the northern entrance, where at 5 m/s the flow was along the wall, while at 10 m/s it is perpendicular to the doors, due to large vortice forming around a neighbouring window. This significantly changes the flow at the northern doors which is not reflected in the averaged value from two openings. Observations formed in the cited literature are consistent with the ones observed in the current study. A change in the performance of natural outlet at different velocities and in differing vortice patterns were studied in isolation in [66] and also observed in [5].

**Table 4.** The difference in mass flow rates at the smoke ventilators at consistent angles and different wind velocities

| ID | Angle | Mass flow rate through vents at 5 m/s | Mass flow rate through vents at 10 m/s | [Mass flow rate through vents at 5 m/s / Mass flow rate through vents at 10 m/s] * 100% | [Avg. velocity at doors at 5 m/s / Avg. velocity at doors at 10 m/s] * 100% |
|---|---|---|---|---|---|
|  | [°] | [kg/s] | [kg/s] | [%] | [%] |
| 1 | 0 | 20.11 | 21.16 | 95% | 99% |
| 2 | 30 | 20.13 | 30.22 | 67% | 87% |
| 3 | 60 | 24.29 | 21.22 | 114% | 100% |
| 4 | 90 | 18.92 | 23.39 | 81% | 112% |
| 5 | 120 | 27.97 | 22.05 | 127% | 122% |
| 6 | 150 | 26.04 | 21.6 | 121% | 115% |
| 7 | 180 | 26.81 | 23.88 | 112% | 109% |
| 8 | 210 | 20.24 | 27.76 | 73% | 97% |
| 9 | 240 | 30.84 | 38.97 | 79% | 90% |
| 10 | 270 | 30.71 | 42.87 | 72% | 81% |
| 11 | 300 | 25.92 | 28.38 | 91% | 98% |
| 12 | 330 | 30.96 | 30.61 | 101% | 96% |

To illustrate the conditions associated with patterns of pressure distribution, flow and pressure field surrounding the building at 5 and 10 m/s wind velocity are presented for following cases (Fig. 5):

- 90° in which we have observed significantly lower system performance compared to the reference, and
- 120° in which we have observed higher system performance at low velocity, rather than at high velocity;
- 270° in which we have observed significantly higher system performance compared to the reference, and
- 330° in which system performed similarly to the reference.

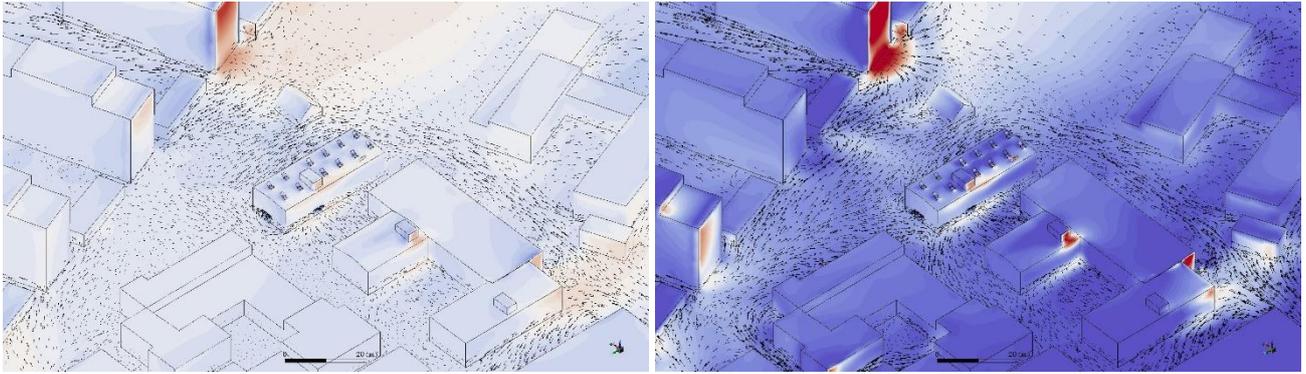
90° (left – 5 m/s, right – 10 m/s) - the cases with the largest (negative) difference between wind and no-wind scenarios

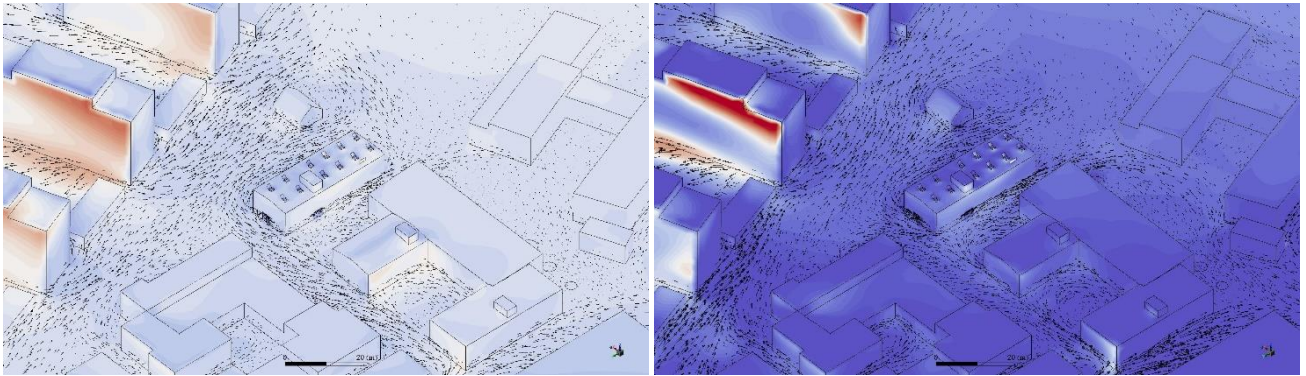
120° (left – 5 m/s, right – 10 m/s) - the cases with better performance at low velocity

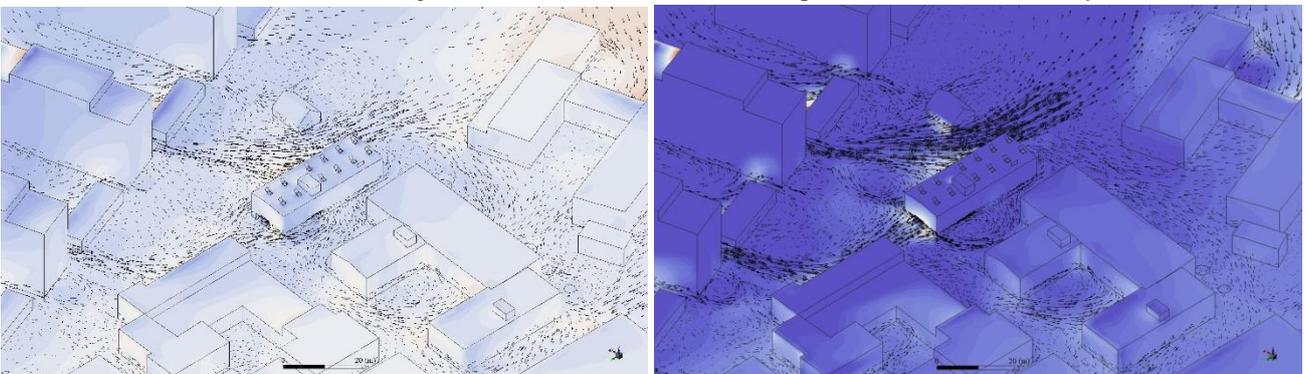
270° (left – 5 m/s, right – 10 m/s) - the cases with the largest (positive) difference between wind and no-wind scenarios

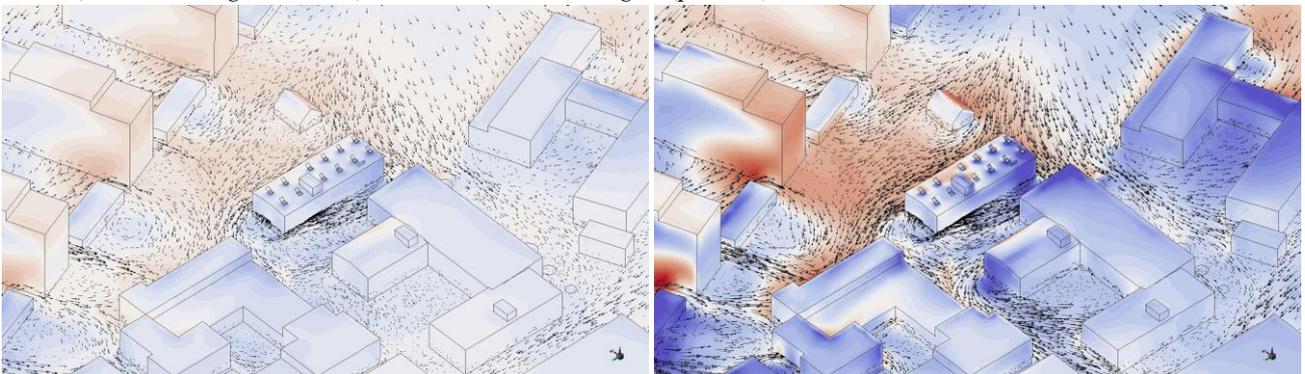
330° (left – 5 m/s, right – 10 m/s) - the cases with small difference between wind and no-wind scenarios

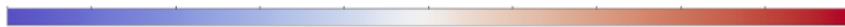

**Figure 5.** Pressure contours(-20…+20 Pa) at the solid boundaries of buildings for 5 m/s and 10 m/s cases with overlayed wind velocity vectors (0.00 – 5.00 m/s) measured at the height of 2.00 m above the ground

The flow inside the warehouse is shown for 0°, 270° and 330° cases (at 10 m/s) and for the reference case without wind in Fig. 6. For cases with positive or neutral effect on the smoke exhaust system's mass flow rate, the flow velocity is generally higher than in no-wind case. For the cases with negative effect on the mass flow rate, the airflow velocity is similar or lower than the reference, although some deviations from that were observed. It was also observed, that the flow from the northern door had larger range and overall impact on the flow inside the compartment, than the flow from the western door even if it had higher velocity. This is in line with previous findings, where the pressure difference between the northern door and outlets was connected to the highest exhaust mass flow rate, while no significant relation was found for the pressure at the western door.

In 0° and 270° cases, the high-velocity flow was observed near the smoke plume, which did disturb the thermal plume. Such a disruption can cause increased air entrainment into the plume, and casue smoke washout from the plume. The consequence is that the plume temperature decreases, and it's buoyancy force are weaker. The washout can be observed in Fig. 5, and is indicated by the distorted (skewed and elongated) shape of the smoke plume, which should be axis-symmetric when undisturbed. However, even though the washout was observed, it did not significantly impact the system performance based on the concentration of the smoke in the warehouse. In the 330° case, high velocity was observed at door 2, but it did not affect the fire. This shows that the fire's spatial location against the inlets may also be a significant factor for the system performance analysis. It is on the contrary to analyses without any wind, in which it was found that the location has not any effects [55].

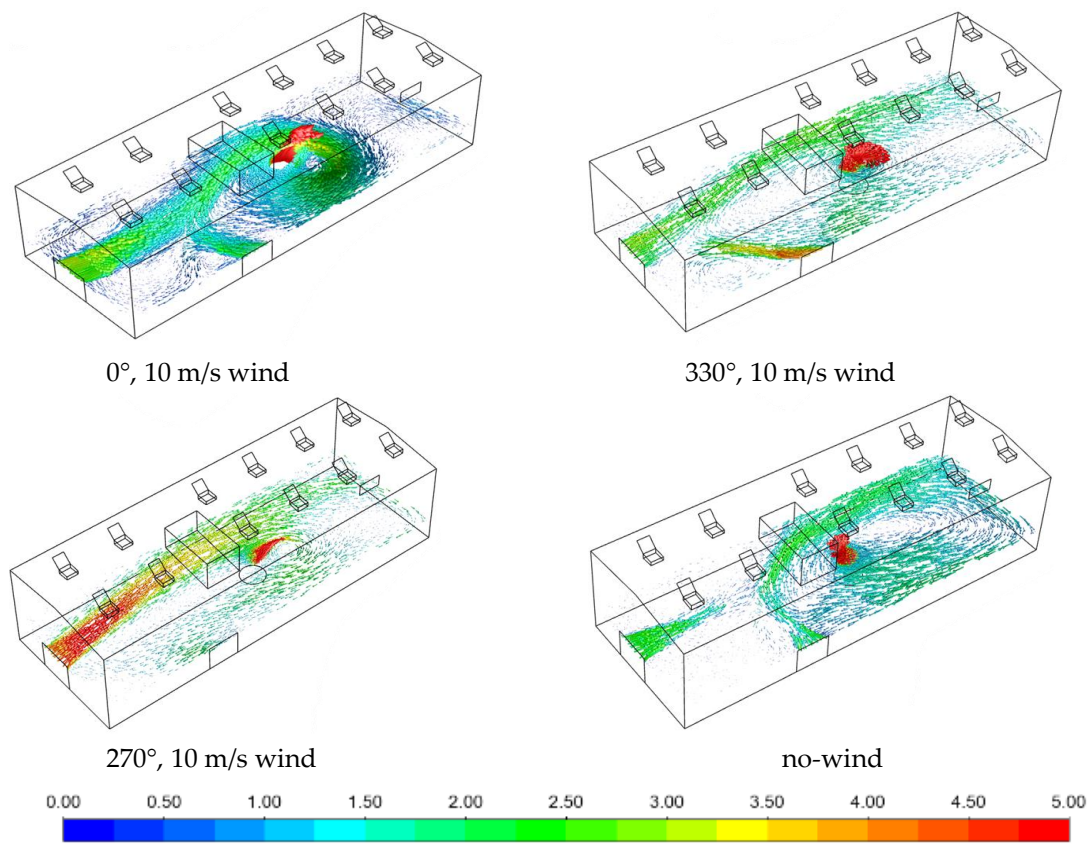

| 0°, 10 m/s wind | 330°, 10 m/s wind |
| 270°, 10 m/s wind | no-wind |

**Figure 6.** Internal flows (0 – 5 m/s) in the warehouse at a height of 2 m above the floor, for 0°, 270° and 330° cases at 10 m/s wind velocity and for the reference case

## 4.3 Pollutant dispersion into the environment

The dispersion of pollutants in the near-field of the warehouse is sensitive to the wind direction and velocity. In order to illustrate the wind effect, the mass density of smoke at the surface of building walls and ground is presented. A logarithmic scale of 0.0001 – 0.1 g/m³ is used in analyses. The upper end of the scale – mass density of smoke equal to 0.1 g/m³ is a value, for which the corresponding visibility of smoke for light-emitting signs is approx. 10 m (4 m for light reflecting signs). This value is often considered as the tenability threshold within buildings in which the fire occurs. For the assessment of the smoke effects outside the building, as described in the introduction, we used the visibility limited ($\tau$ = 2%) to 100 m and 200 m, which corresponded to the smoke density equal to 0.0045 g/m³ and 0.00223 g/m³, respectively. The lower end of the scale 0.001 – 0.0001 g/m³ represents conditions at which the smoke is hardly detectable (concentration of approx. 1 – 0.1 ppm). When investigating the dispersion effects, it must be noted that the simulated fire is of moderate size in present study and represents the lower end of the industrial fires' range. Considering this we would like to emphasise the complexity of urban flows and their impact on the smoke dispersion in the neighbourhood, rather than the acute effects of fire (which are correlated with the size of the fire). Results of the simulations showing the smoke dispersion in the near-field for all wind angles and 5 and 10 m/s wind velocities, are shown in figures A1 and A2 in Appendix A.

Cases 30° – 90° of wind angle show instances of the downwash effect on smoke behind a group of tall buildings, that form the eastern boundary of the urban street canyon. The vortices behind these buildings lead to the formation of large areas with the high density of smoke, at a considerable distance from the fire (over 200 m). This effect is similar for both wind velocities, although the smoke plume is wider in case of strong wind. The occupancy of these buildings is residential and the evacuation of the occupants would be justified in analysed scenarios. It is worth noting that the tenants of buildings located closer to the fire are exposed to lower concentrations of combustion products, than the occupants of the buildings behind which these large vortices were observed.

The urban street canyon effect is well visible in case 120° of the wind angle and is very similar for both wind velocities. The smoke is constrained in the area between two rows of high buildings, and the flow direction is not aligned with the wind inflow. Some up-wind smoke movement is also observed for these scenarios and is more clearly visible in strong wind conditions. In analysed case, the evacuation of buildings along the street may be necessary. Furthermore, due to the significant reduction of visibility in the street over its whole width, and at a length of over 200 m, the street would have to be closed for the road traffic. Otherwise, the smoke obscuration in this area would affect the driving abilities [67] and significantly increase the risk of traffic accidents [68].

In some scenarios (0°, 60°, 150°, 180°, 240°, 300°) the shape of the smoke-exposed area is narrow, which may indicate that the smoke plume was narrow and likely high above the ground. In other cases (90°, 210°) the exposed area is wide, or very wide (330°). To investigate the shape of the smoke plumes three-dimensional iso-surface plots of smoke mass density (0.001 g/m³, which corresponds to approx. 1 ppm of soot) were created, as shown in Fig. 7. In case of the very wide plume (330°), the area with a high concentration of pollutants spans over multiple neighbouring buildings. In terms of environmental pollution, this case would result in the most significant near-field contamination. The local concentration of smoke in the warehouse building's close neighbourhood is substantially higher than in other cases.

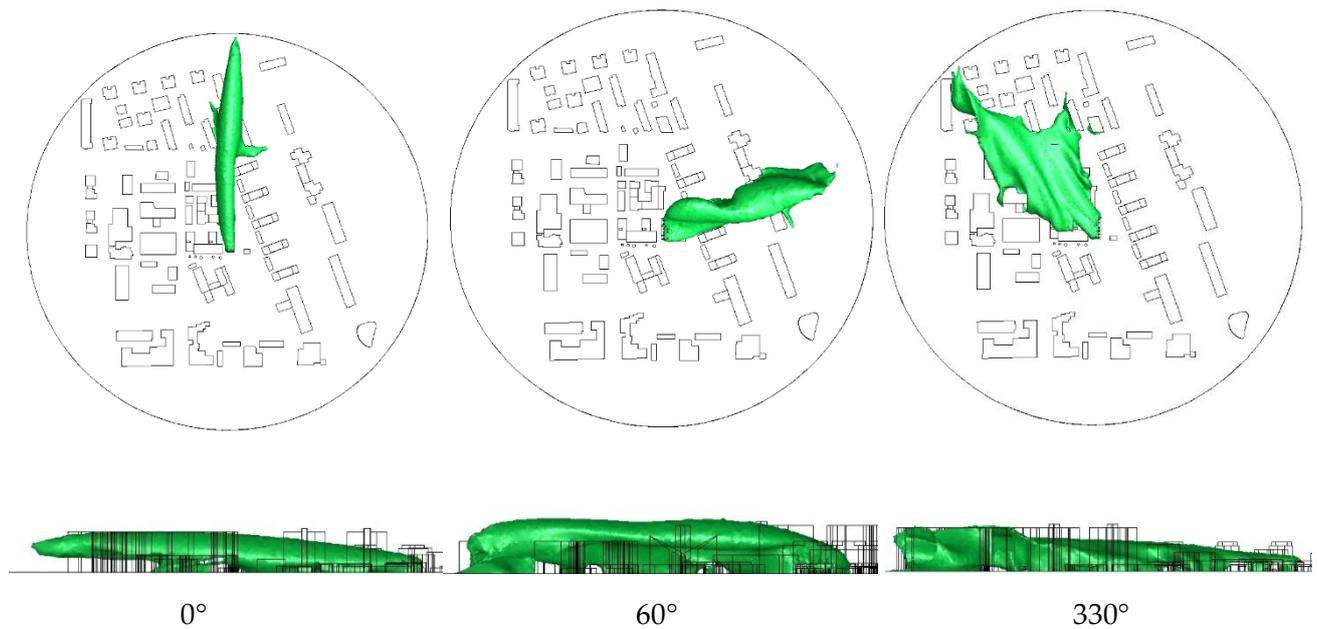

0°      60°      330°

**Figure 7.** Iso-surface plots of the smoke mass density (0.001 g/m³) highlighting different plume shapes at $u_{ref}$ = 10 m/s. Narrow plume (0°), wide plume (210°) and very wide plume (330°)

Fig. 8 presents the smoke concentration at the ground and walls of buildings close to the investigated warehouse. In most cases, the downwind buildings are exposed to smoke concentrations between 0.005 and 0.01 g/m³. However, in the case of 330° wind angle and 10 m/s velocity, the smoke is pushed on the group of buildings with heights similar to the warehouse's height. In this case, the smoke gathers in court areas of buildings and in the narrow passages between them. Smoke contamination above 0.015 g/m³ (approx. 30 m of visibility range) and above 0.0045 g/m³ was observed at the distances of 130 m and 250 m from the fire, respectively. These local effects were larger for the strong wind. This may be attributed to more significant building wake area and downwash effects at high wind velocities.

The observed diversity of results related to smoke dispersion in the building near-field illustrates the complexity of mitigating the consequences of fires in dense, urban environments. With the use of CFD modelling, most onerous scenarios can be identified and potentially used for civil preparedness studies. Furthermore, post-fire analyses may be performed to identify the potential locations of smoke contamination, which, may be located far away from the source of the fire. In case of far field investigations, other tools may be more applicable due to computational requirements, compared to the CFD models [11].

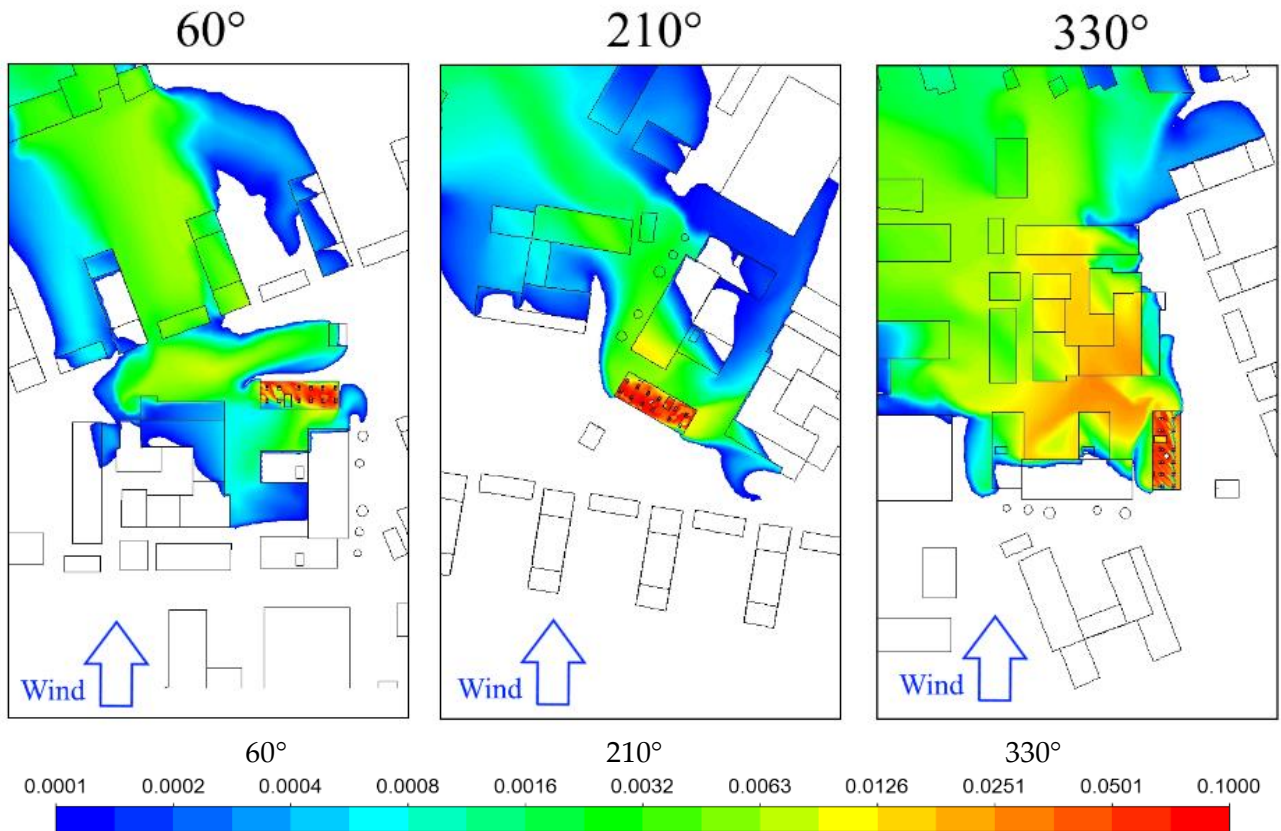

**Figure 8.** The smoke mass density in the near-field, at the surface of buildings and ground (0.0001 – 0.1 g/m³, log scale). Wind angles 60°, 210° and 330°, $u_{ref}$ = 10 m/s

## 5   Conclusions

The influence of neighbouring architecture was included in a parametric investigation on the wind effects on a building fire and smoke exhaust performance. This was done by performing 25 CFD simulations for two different wind velocities and twelve different wind angles, and for the reference case without wind. The simulations were not validated with experiments, due to costs associated with carrying full scale wind-fire experiments and ethical constrains related to releasing fire pollutants in densely populated urban environment. To reduce the modelling uncertainties, the research follows the existing body of literature in which CFD (ANSYS Fluent in particular) was used to study urban dispersion of pollutants (as shown in Chapter 2.3), and strictly followed both Computational Wind Engineering and wind and fire coupled modelling best practice guidelines and recommendations (Chapter 3.3). Having this in mind, the CFD modelling allowed to investigate the impact of surrounding buildings on the wind field and the outcomes of the fire inside and outside the investigated warehouse.

The wind velocity affected smoke removal from the building and its further dispersion in the environment. Based on the simulations' variability, it is not possible to claim that higher velocity is always associated with worse outcomes. In half of the tested scenarios, the smoke exhaust performance was better in strong, rather than in moderate wind. In 25% of cases, the ventilation performance in wind conditions exceeded that estimate without the wind. These observations are consistent with some previous studies reported in the literature and in [2]. Differences were observed with regards to both the direction and velocity of the wind, and associated to wind patterns in the building vicinity. The smoke ventilators' measured mass flow rate varied between 74%-114% and 78%-158% of the reference mass flow at the moderate and strong wind, respectively. The highest ventilation system performance was associated with the wind causing strong flow through the northern entrance into the warehouse, also associated with the highest pressure difference between the doors and the outlets (8.4 – 11.9 Pa). The lower system performance was generally associated with the lower pressure difference between the doors and the outlets. For four of the modelled wind directions (33%), the system perfomance was higher in the moderate rather than in strong wind conditions, although in most of these cases this performance was still worse than in no-wind conditions. These observations show, that the choice of the most onerous scenarios for wind and fire coupled analyses could be potentially determined through steady-state 'cold flow' analysis of pressure differences at builing roof and facades. This could be used to considerably reduce the number of transient fire scenarios to be investigated, reducing the workload required to understand the impact of wind on natrual ventilation of buildings in their design process.

Outside of the modelled warehouse the wind had a significant impact on pollutant transport. Even though the fire was limited to the moderate size of 8.00 MW, the observed pollutant concentrations in the building near-field were considerable, causing visibility to be limited below 30 m in the area spanning 115 m from the fire (330°, 10 m/s). Furthermore, a significant amount of smoke was observed behind tall buildings located up to 250 m from the fire at 30° and 90° cases. This smoke accumulation was associated with strong vortices forming behind these buildings, that trapped the smoke. In fact, the smoke concentration behind these buildings was higher, than in areas around other buildings closer to the fire, and could be potentially dangerous at prolonged exposure.

The CFD simulations have shown, that depending on the wind angle, and the neighbouring architecture, the shape of thermal smoke plume changed from almost the ideal thermal plume (at 0°) to flat and wide dispersion over a large area (at 330°). In the latter case, the smoke has accumulated in the driveways and courtyards of neighbouring low buildings. We have also observed the effects of a nearby urban street canyon (at 120°), where the smoke was transported between two rows of buildings (also upstream the wind). The street area was filled with smoke of the concentration that limited visibility below 100 m. Such conditions may increase risk related to the road traffic and may require to close a part of the city road network.

Fires in the built environment may be significantly larger than the one reported in this case study, with their HRR measured in hundreds of megawatts. This means the resulting smoke concentrations and the contaminated area may be considerably higher and larger than in the analysed scenarios. This also means that the wind and fire coupled modelling may and should be used to investigate the outcomes of such fires, in order to improve the civil preparedness and public response to them. Finally, research performed after environmentally significant fires may help to understand the actions required to mitigate the fire consequences and its environmental impact.

**Funding**

This work was funded by the Ministry of Science and Higher Education of Poland, through the Building Research Institute statutory funding grant NZP-114/2020. Parts of the research performed in the revision were funded by the National Science Centre, Poland, on the basis of a contract for the implementation and financing of a research project No 2020/37/B/ST8/03839.

**References**

[1]     T. Tanaka, Vent Flows, in: SFPE Handb. Fire Prot. Eng., Springer New York, New York, NY, 2016: pp. 455–485. https://doi.org/10.1007/978-1-4939-2565-0_15.

[2]     W. Węgrzyński, T. Lipecki, Wind and Fire Coupled Modelling—Part I: Literature Review, Fire Technol. 54 (2018) 1405–1442. https://doi.org/10.1007/s10694-018-0748-5.

[3]     R.N. Meroney, Wind effects on atria fires, J. Wind Eng. Ind. Aerodyn. 99 (2011) 443–447. https://doi.org/10.1016/j.jweia.2010.11.003.

[4]     W. Węgrzyński, G. Krajewski, Influence of wind on natural smoke and heat exhaust system performance in fire conditions, J. Wind Eng. Ind. Aerodyn. 164 (2017) 44–53. https://doi.org/10.1016/j.jweia.2017.01.014.

[5]     W. Węgrzyński, G. Krajewski, P. Suchy, T. Lipecki, The influence of roof obstacles on the performance of natural smoke ventilators in wind conditions, J. Wind Eng. Ind. Aerodyn. 189 (2019) 266–275. https://doi.org/10.1016/j.jweia.2019.04.004.

[6]     M. Król, A. Król, Multi-criteria numerical analysis of factors influencing the efficiency of natural smoke venting of atria, J. Wind Eng. Ind. Aerodyn. 170 (2017) 149–161. https://doi.org/10.1016/j.jweia.2017.08.012.

[7]     A.A. Stec, K. Dickens, J.L.J. Barnes, C. Bedford, Environmental contamination following the Grenfell Tower fire, Chemosphere. 226 (2019) 576–586. https://doi.org/10.1016/j.chemosphere.2019.03.153.

[8]     E. Peltier, J. Glanz, W. Cai, J. White, Notre-Dame's Toxic Fallout, New York Times. (2019). https://www.nytimes.com/interactive/2019/09/14/world/europe/notre-dame-fire-lead.html.

[9]     M. Mcnamee, G. Marlair, B. Truchot, B.J. Meacham, Research Roadmap : Environmental Impact of Fires in the Built Environment, 2020.

[10]    K. Himoto, T. Tanaka, Development and validation of a physics-based urban fire spread model, Fire Saf. J. 43 (2008) 477–494. https://doi.org/10.1016/j.firesaf.2007.12.008.

[11]    W. Węgrzyński, T. Lipecki, Fire and Smoke Modelling, in: Handb. Fire Environ., 2023: pp. 101–181. https://doi.org/10.1007/978-3-030-94356-1_4.

[12]    P.H. Thomas, P.L. Hinkley, C.R. Theobald, D.L. Simms, Investigations into the flow of hot gases in roof venting, Her Majesty's Stationary Office, London, 1963.

[13]    NFPA, NFPA 204 Standard for Smoke and Heat Venting 2015 Edition, (2015).

[14]    M. Poreh, S. Trebukov, Wind effects on smoke motion in buildings, Fire Saf. J. 35 (2000) 257–273. https://doi.org/10.1016/S0379-7112(00)00017-5.

[15]    L. Moosavi, N. Mahyuddin, N. Ab Ghafar, M. Azzam Ismail, Thermal performance of atria: An overview of natural ventilation effective designs, Renew. Sustain. Energy Rev. 34 (2014) 654–670.

https://doi.org/10.1016/j.rser.2014.02.035.

[16] ISO TC92, ISO/TR 24188:2022(en) Large outdoor fires and the built environment — Global overview of different approaches to standardization, 2022.

[17] H. Ingason, B. Persson, Effects of Wind on Natural Fire Vents, in: Brand. Proj. 055-921; SP Rep. 199504, SP Swedish National Testing and Research Institute, Fire Technology, 1995.

[18] W. Węgrzyński, G. Krajewski, Wind influence in numerical analysis of NSHEVs performance, in: 2nd Fire Evacuation Model. Tech. Conf., Torremolinos, Spain, Spain, 2016.

[19] M. Król, Numerical studies on the wind effects on natural smoke venting of atria, Int. J. Vent. 15 (2016) 67–78. https://doi.org/10.1080/14733315.2016.1173293.

[20] H. Li, C. Fan, J. Wang, Effects of Wind and Adjacent High Rise on Natural Smoke Extraction in an Atrium with Pitched Roof, APCBEE Procedia. 9 (2014) 296–301. https://doi.org/10.1016/j.apcbee.2014.01.053.

[21] W.K. Chow, P. Liu, G.W. Zou, Wind Effect on Smoke Exhaust Systems in a Large Cargo Hall with Two Compartments, in: Cairns, Australia, 2007: pp. 1415–1422.

[22] S. Kerber, D. Madrzykowski, Fire fighting tactics under wind driven conditions: 7-story building experiments, in: NIST Tech. Note 1629, National Institute of Standards and Technology, Gaithersburg, MD, 2009.

[23] D. Madrzykowski, S. Kerber, Fire fighting tactics under wind driven conditions: laboratory experiments, in: NIST Tech. Note 1618, National Institute of Standards and Technology, Gaithersburg, MD, 2009.

[24] A. Barowy, D. Madrzykowski, Simulation of the Dynamics of a Wind-Driven Fire in a Ranch-Style House - Texas, 2012.

[25] C. Weinschenk, C. Beal, O.A. Ezekoye, Modeling fan-driven flows for firefighting tactics using simple analytical models and CFD, J. Fire Prot. Eng. 21 (2011) 85–114. https://doi.org/10.1177/1042391510395694.

[26] S.L. Manzello, R. Blanchi, M.J. Gollner, D. Gorham, S. McAllister, E. Pastor, E. Planas, P. Reszka, S. Suzuki, Summary of workshop large outdoor fires and the built environment, in: Fire Saf. J., Elsevier Ltd, 2018: pp. 76–92. https://doi.org/10.1016/j.firesaf.2018.07.002.

[27] S.L. Manzello, K. Almand, E. Guillaume, S. Vallerent, S. Hameury, T. Hakkarainen, FORUM position paper, Fire Saf. J. 100 (2018) 64–66. https://doi.org/10.1016/j.firesaf.2018.07.003.

[28] R.J. McDermott, N.P. Bryner, J.A. Heintz, Large Outdoor Fire Modeling (LOFM) workshop summary report, Gaithersburg, MD, 2019. https://doi.org/10.6028/NIST.SP.1245.

[29] S.E. Caton, R.S.P. Hakes, D.J. Gorham, A. Zhou, M.J. Gollner, Review of Pathways for Building Fire Spread in the Wildland Urban Interface Part I: Exposure Conditions, Fire Technol. 53 (2017) 429–473. https://doi.org/10.1007/s10694-016-0589-z.

[30] A. Tohidi, N.B. Kaye, Stochastic modeling of firebrand shower scenarios, Fire Saf. J. 91 (2017) 91–102. https://doi.org/10.1016/j.firesaf.2017.04.039.

[31] A. Leelossy, F. Molnar Jr., F. Izsak, A. Havasi, I. Lagzi, R. Meszaros, Dispersion modeling of air pollutants in the atmosphere : a review, Cent. Eur. J. Geosci. 6 (2014) 257–278. https://doi.org/10.2478/s13533-012-0188-6.

[32] C.C. Simpson, J.J. Sharples, J.P. Evans, Resolving vorticity-driven lateral fire spread using the WRF-Fire coupled atmosphere-fire numerical model, Nat. Hazards Earth Syst. Sci. 14 (2014) 2359–2371. https://doi.org/10.5194/nhess-14-2359-2014.

[33] N.H. Al-Khalidy, Utilising a Combination of Computational Fluid Dynamics and Standard Air Quality Simulation, Int. J. Mech. 11 (2017) 210–217.

[34] P. Gousseau, B. Blocken, T. Stathopoulos, G.J.F. van Heijst, CFD simulation of near-field pollutant dispersion on a high-resolution grid: A case study by LES and RANS for a building group in downtown Montreal, Atmos. Environ. 45 (2011) 428–438. https://doi.org/10.1016/j.atmosenv.2010.09.065.

[35] T. Stathopoulos, L. Louis, P. Saathoff, A. Gupta, The effect of stack height, stack location and rooftop structures on air intake contamination, Stud. Res. Proj. (2004).

[36] J. Franke, A. Hellsten, H. Schlünzen, B. Carissimo, Best practice guideline for the CFD simulation of flows in the urban environment, COST Office Brussels, 2007.

[37] B. Blocken, R. Vervoort, T. van Hooff, Reduction of outdoor particulate matter concentrations by local removal in semi-enclosed parking garages: A preliminary case study for Eindhoven city center, J. Wind Eng. Ind. Aerodyn. 159 (2016) 80–98. https://doi.org/10.1016/j.jweia.2016.10.008.

[38] N. Antoniou, H. Montazeri, M. Neophytou, B. Blocken, CFD simulation of urban microclimate: Validation using high-resolution field measurements, Sci. Total Environ. 695 (2019) 133743. https://doi.org/10.1016/j.scitotenv.2019.133743.

[39] S. Liu, W. Pan, H. Zhang, X. Cheng, Z. Long, Q. Chen, CFD simulations of wind distribution in an urban community with a full-scale geometrical model, Build. Environ. 117 (2017) 11–23. https://doi.org/10.1016/j.buildenv.2017.02.021.

[40] S. Liu, W. Pan, X. Zhao, H. Zhang, X. Cheng, Z. Long, Q. Chen, Influence of surrounding buildings on wind flow around a building predicted by CFD simulations, Build. Environ. 140 (2018) 1–10. https://doi.org/10.1016/j.buildenv.2018.05.011.

[41] L. Brzozowska, Modelling the propagation of smoke from a tanker fire in a built-up area, Sci. Total Environ. 472 (2014) 901–911. https://doi.org/10.1016/j.scitotenv.2013.11.130.

[42] T. Nozu, T. Tamura, LES of turbulent wind and gas dispersion in a city, J. Wind Eng. Ind. Aerodyn. 104–106 (2012) 492–499. https://doi.org/10.1016/j.jweia.2012.02.024.

[43] BSI, The application of fire safety engineering principles to fire safety design of buildings - Part 6: Human factors: Life safety strategies - Occupant evacuation, behavious and condition (Sub-system 6), PD 7974-6. (2004).

[44] G. Sztarbała, An estimation of conditions inside construction works during a fire with the use of Computational Fluid Dynamics, Bull. Polish Acad. Sci. Tech. Sci. 61 (2013) 155–160. https://doi.org/10.2478/bpasts-2013-0014.

[45] D.A. Purser, Combustion Toxicity, in: SFPE Handb. Fire Prot. Eng., Springer New York, New York, NY, 2016: pp. 2207–2307. https://doi.org/10.1007/978-1-4939-2565-0_62.

[46] A.A. Stec, Fire toxicity – The elephant in the room?, Fire Saf. J. (2017). https://doi.org/10.1016/j.firesaf.2017.05.003.

[47] Q. Zhang, Image based analysis of visibility in smoke laden environments, (2010).

[48] G.W. Mulholland, C. Croarkin, Specific extinction coefficient of flame generated smoke, Fire Mater. 24 (2000) 227–230. https://doi.org/10.1002/1099-1018(200009/10)24:5<227::AID-FAM742>3.0.CO;2-9.

[49] W. Węgrzyński, T. Lipecki, G. Krajewski, Wind and Fire Coupled Modelling—Part II: Good Practice Guidelines, Fire Technol. 54 (2018) 1443–1485. https://doi.org/10.1007/s10694-018-0749-4.

[50] ANSYS, ANSYS Fluent 14.5.0 - Technical Documentation, 2014.

[51] CEN, EN 12101-2 Smoke and heat control systems Part 2: Specification for natural smoke and heat exhaust ventilators, (2015).

[52] S. Liu, W. Pan, H. Zhang, X. Cheng, Z. Long, Q. Chen, CFD simulations of wind distribution in an urban community with a full-scale geometrical model, Build. Environ. 117 (2017) 11–23. https://doi.org/10.1016/j.buildenv.2017.02.021.

[53] K. McGrattan, S. Hostikka, R. McDermott, J. Floyd, C. Weinschenk, K. Overholt, Fire Dynamics Simulator User's Guide, Sixth Edition, 2017. https://doi.org/10.6028/NIST.SP.1019.

[54] K. Hill, J. Dreisbach, F. Joglar, B. Najafi, K. McGrattan, R. Peacock, A. Hamins, Verification and Validation of Selected Fire Models for Nuclear Power Plant aplications, volume 7, Washington DC, 2014.

[55] W. Węgrzyński, M. Konecki, Influence of the fire location and the size of a compartment on the heat and smoke flow out of the compartment, AIP Conf. Proc. 1922 (2018) 110007. https://doi.org/10.1063/1.5019110.

[56] W. Węgrzyński, G. Vigne, Experimental and numerical evaluation of the influence of the soot yield on the visibility in smoke in CFD analysis, Fire Saf. J. 91 (2017) 389–398. https://doi.org/10.1016/j.firesaf.2017.03.053.

[57] F.R. Menter, Two-equation eddy-viscosity turbulence models for engineering applications, AIAA J. 32 (1994) 1598–1605. https://doi.org/10.2514/3.12149.

[58] P. Tofiło, R. Porowski, W. Węgrzyński, Spatial distribution of thermal radiation – Verification of the finite volume method, in: Interflam, 2016.

[59] Y. Toparlar, B. Blocken, B. Maiheu, G.J.F. van Heijst, A review on the CFD analysis of urban microclimate, Renew. Sustain. Energy Rev. 80 (2017) 1613–1640. https://doi.org/10.1016/j.rser.2017.05.248.

[60] N. Antoniou, H. Montazeri, H. Wigo, M.K.A. Neophytou, B. Blocken, M. Sandberg, CFD and wind-tunnel analysis of outdoor ventilation in a real compact heterogeneous urban area: Evaluation using "air delay," Build. Environ. 126 (2017) 355–372. https://doi.org/10.1016/j.buildenv.2017.10.013.

[61] M. Lateb, R.N. Meroney, M. Yataghene, H. Fellouah, F. Saleh, M.C. Boufadel, On the use of numerical modelling for near-field pollutant dispersion in urban environments - A review, Environ. Pollut. 208 (2016) 271–283. https://doi.org/10.1016/j.envpol.2015.07.039.

[62] R. Meroney, R. Ohba, B. Leitl, H. Kondo, D. Grawe, Y. Tominaga, Review of CFD Guidelines for Dispersion Modeling, Fluids. 1 (2016) 14. https://doi.org/10.3390/fluids1020014.

[63] K. Luo, H.J. Yu, Z. Dai, M.M. Fang, J. Fan, CFD simulations of flow and dust dispersion in a realistic urban area, Eng. Appl. Comput. Fluid Mech. 10 (2016) 229–243. https://doi.org/10.1080/19942060.2016.1150205.

[64] B. Blocken, 50 years of Computational Wind Engineering: Past, present and future, J. Wind Eng. Ind. Aerodyn. 129 (2014) 69–102. https://doi.org/10.1016/j.jweia.2014.03.008.

[65] B. Blocken, LES over RANS in building simulation for outdoor and indoor applications: A foregone conclusion?, 2018. https://doi.org/10.1007/s12273-018-0459-3.

[66] W. Węgrzyński, G. Krajewski, G. Kimbar, A concept of external aerodynamic elements in improving the performance of natural smoke ventilation in wind conditions, in: AIP Conf. Proc., 2018: p. 110006. https://doi.org/10.1063/1.5019109.

[67] N. Wetterberg, E. Ronchi, J. Wahlqvist, Individual Driving Behaviour in Wildfire Smoke, Fire Technol. (2020). https://doi.org/10.1007/s10694-020-01026-5.

[68] X. Cheng, B. Yang, G. Liu, T. Olofsson, H. Li, A variational approach to atmospheric visibility estimation in the weather of fog and haze, Sustain. Cities Soc. 39 (2018) 215–224. https://doi.org/10.1016/j.scs.2018.02.001.